\documentclass[UTF8,a4paper]{article}
\usepackage{amsmath}
\usepackage{graphicx}
\usepackage{subfigure}
\usepackage{epstopdf}
\usepackage{geometry}
\usepackage{ulem}
\usepackage{indentfirst}
\usepackage{amsfonts,amssymb,dsfont}
\usepackage{setspace}
\usepackage{threeparttable}
\usepackage{amsmath,mathtools,amsthm}
\usepackage{array}
\usepackage{extarrows}
\usepackage{upgreek}

\makeatletter
\renewcommand*\env@matrix[1][\arraystretch]{%
  \edef\arraystretch{#1}%
  \hskip -\arraycolsep
  \let\@ifnextchar\new@ifnextchar
  \array{*\c@MaxMatrixCols c}}
\makeatother
\begin{document}


\title{\bf Dynamical polarization, plasmon model, and the Friedel oscillation of the screened potential in doped
Dirac and Weyl system}
\author{Chen-Huan Wu
\thanks{chenhuanwu1@gmail.com}
\\Key Laboratory of Atomic $\&$ Molecular Physics and Functional Materials of Gansu Province,
\\College of Physics and Electronic Engineering, Northwest Normal University, Lanzhou 730070, China}

\maketitle
\vspace{-30pt}
\begin{abstract}
\begin{large} 

We discuss the dynamical polarization, plasmon dispersion, relaxation time,
and the Friedel oscillation of screened potential 
of the two-dimension Dirac and three-dimension Weyl system (which are gapped) in the low-energy tigh-binding model.
The results, like the Fermi wavevector, Thomas-Fermi wavevector, and longitudinal conductivity are obtained in different
dimensions.
Some important conclusions are detailedly discussed in this paper, including the 
screening character under short or long range Coulomb interaction,
and the longitudinal conductivity in two- or three-dimensions.
The longitudinal conductivity in optical limit is distinguishing for the case of two-dimension system and three-dimension system.
The density-dependence (including the carrier density and the impurity concentration) of the 
Fermi wavevector, dc conductivity, and the relaxation time are discussed.
Specially, for the doped Weyl system, the pumped carrier density due to the chiral anomaly origin from electromagnetic response
is controlled by the internode relaxation time which has also been analyzed.
Our results is helpful to the application of the Dirac or Weyl systems as well as the study on their low-temperature characters.\\




\end{large}

\end{abstract}
\begin{large}

\section{Introduction}

In this paper, we discuss the dynamical polarization, plasmon model, screened potential of the charged impurities, and the relaxation time
in the presence of the impurity of the 
two-dimension Dirac materials (like the silicene and MoS$_{2}$ which both with hexagonal and buckled 
lattice structure and
strong spin-orbit coupling (SOC)) and the three-dimension Weyl material.
The dynamical polarization is a renormalization of the Coulomb interaction between carriers\cite{Malcolm J D},
which is helpful during the study of screened properties as well as the collective excitation models.
In Weyl semimetal (gapless), the dominating scattering at wave vector ${\bf q}=2{\bf k}_{F}$ in momentum space
induces the Friedel oscillation of the density of states or the screened potential of the charged impurites,
however, in Weyl nodes the chiral anomaly (i.e., ${\bf E}\cdot{\bf B}\neq 0$ as a result of the nontrivial topology) 
suppress the backscattering at ${\bf q}=2{\bf k}_{F}$,
that thus leads to the continuous first derivative but discontinuous second derivative of polarization ($(d^{2}/d^{2}{\bf q})\Pi({\bf q},\omega)$),
that's similar to the gapless graphene or silicene 
which have discontinuous first derivative of polarization when gapped,
thus we can speculate that the gapped Weyl system also has discontinuous first derivative of polarization.
We also found that
the kink of static polarization in ${\bf q}=2{\bf k}_{F}$ is more obvious in the silicene\cite{Wu C H3,XX,Wu C H_3}, MoS$_{2}$ or graphene
than that in the Weyl semimetal or two-dimension pseudospin-1 dice lattice\cite{Malcolm J D}.
The suppression of the backscattering also leads to the faster decay of the screened Coulomb potential of the charged impurity,
which is found decays as $\sim{\rm sin}(2{\bf k}_{F}r)/r^{4}$ during the intranode process in Weyl semimetal\cite{Lv M}
(such strong decay also found in the dice lattice which acts as $\sim{\rm cos}(2{\bf k}_{F}r)/r^{4}$\cite{Malcolm J D})
with broken time-reversal invariance,
that's 
much faster than the $\sim{\rm sin}(2{\bf k}_{F}r)/r^{2}$ or $\sim{\rm cos}(2{\bf k}_{F}r)/r^{3}$ as found in graphene, 
silicene\cite{Wu C H3,XX,Wu C H_3,Chang H R} or 
two-dimension electron gas (2DEG)\cite{Wu C H3,Wunsch B}.
In two- or three-dimension Dirac system (including the topological insulators
but not including the 2DEG),
the existent quasiparticle chirality also suppress the backscattering,
thus we can suspect such fast decay ($\sim r^{-4}$) also exist in Dirac system with quasiparticle chirality or warping structure,
but we note here that 
the elastic intervalley scattering is benefit to restoration of the suppressed quasiparticle interference (also by the hexagonal warping\cite{XX}) 
and the backscattering,
such restoration of backscattering can also origin from the internodal scattering in three-dimension Dirac or Weyl semimetal\cite{Lv M}.
The internodal scattering between the Weyl nodes with opposite chirality
related to the internode charge relaxation time which is much shorter than the intervalley scattering one.
The internode charge relaxation time 
(which can be estimated by the mean-free path $\ell_{3}=v_{F}\tau\sim 1/\sqrt[3]{n_{i}}$
with $n_{i}$ is the impurity concentration) in Weyl semimetal is shorter than that in the Dirac semimetal, 
e.g., for the Weyl semimetal Eu$_{2}$Ir$_{2}$O$_{7}$ which with Fermi velocity $v_{F}=4\times 10^{5}$ m/s
and slower than in the silicene (whose Fermi velocity is $v_{F}=5.5\times 10^{5}$ m/s), 
the internode charge relaxation time is nearly 25 fs 
which is much longer than that in silicene (nearly 18.2 ps).
The chirality effect of the Dirac materials can be shown in the interband or intraband scattering matrix element
which describe the overlap of the eigenstates
as shown in the noninteracting (inreducible) or interacting expression of the dynamical polarization\cite{Wu C H3,XX,Sensarma R,Sarma S D}.
For nonchiral systems, like the conventional 2D electron systems, the scattering matrix element equals one.

For two-dimension Dirac system in this paper (where we mainly discuss the silicene), we apply the dielectric constant of the surrounding matter as air/SiO$_{2}$
with $\epsilon=(3.9+1)/2=2.45$.
The effective dielectric constant for silicene-like Dirac two-dimension system 
reads $\epsilon^{*}=1+g_{s}g_{v}\pi r_{s}/8$\cite{Hwang E H} in large ${\bf q}$ or $\mu$,
where $g_{s}g_{v}$ denotes the spin and valley degenerate here and the Wigner-Seitz radiu $r_{s}=e^{2}/\hbar\epsilon\gamma$ for the monolayer silicene
(or graphene)
with $\gamma$ the band parameter.
While for bilayer silicene or other bilayer system with both the interlayer and intralayer hopping, 
$r_{s}=e^{2}m^{*}/\epsilon\epsilon^{*}{\bf k}_{F}$
with the effective mass $m^{*}=(t^{{\rm inter}})^{2}/2 v_{F}^{2}=\hbar^{2}t^{{\rm inter}}/(2t^{2})\sim 1/v_{F}^{2}$
where $t^{{\rm inter}}$ is the strength of the interlayer hopping.
$\gamma=2{\bf k}_{F}/(\pi\Pi({\bf q},0)\hbar)$ ($\gamma\sim v_{F}$ for the monolayer silicene or graphene with large carriers density) 
is inversely proportional to the static polarization function\cite{Wu C H3}.
Such effective background dielectric constant
is larger than the original one, but for bilayer silicene or the 2DEG, $\epsilon^{*}=1$\cite{Hwang E H2}
due to the vanishing contribution of the $\Pi^{-}({\bf q},\omega)$\cite{Lv M,Sensarma R}.
Thus we can know that the $\epsilon^{*}$ is proportion to the polarization function
and it's also proportional to the interband transition as well as the longitudinal conductivity.
The $\epsilon^{*}$ is measured as 2.09 for graphene on SiO$_{2}$ insulator substrate\cite{Sodemann I},
and measured as 4.4\cite{Pyatkovskiy P K} for the freestanding graphene with zero-magnetic field.

The $r_{s}$ is usually close to 2.16 for the freestanding monolayer silicene or graphene
but larger than 10 for the freestanding bilayer\cite{Sensarma R} or Hydrogenated ones\cite{Guillemette J}.
while for the monolayer graphene on SiO$_{2}$, the Wigner-Seitz radius $r_{s}=2.16/3.9\approx 0.55$,
close to the result reported in Refs.\cite{Hwang E H,Malcolm J D}.
The bilayer silicene or graphene, especially when with a large impurity concentration ($n\sim 10^{12}$ cm$^{-2}$),
$r_{s}\gg 1$ and even $>10$ acts like the the Wigner crystal.
The effective dielectric constant is important especially in the short-wavelength case with ${\bf q}>2{\bf k}_{F}$,
where the screened Coulomb interaction well obey the expression of the Fourier transformation of the two-dimension Coulomb interaction 
$2\pi e^{2}/\epsilon\epsilon^{*}{\bf q}$ as shown in our previous result (see Fig.5(a) of Ref.\cite{Wu C H_3}).
For the impurity medium, the longitudinal conductivity and effective dielectric function as well as the permeability are
also related to the position and the two-point correlation fucntion.
The static effective dielectric constant is obtained by Ref.\cite{Sodemann I}
and the static effective dielectric constant without the interband transition is obtained by Ref.\cite{Keller J B}.

\section{Models}

In both the Dirac or Weyl semimetal, the Berry curvature together with the induced anomalous velocity term 
requires either the broken time-reversal invariance or the inversion symmetry\cite{Wu S}.
For three dimension Dirac or Weyl semimetal,
the Berry gauge field can be origin from the magnetic monopole (e.g., the Dirac or Weyl nodes or the node line) 
which carries the monopole charge $2s$\cite{Wu C HX},
such phenomeno can also be found in the multi-band touching system\cite{Ezawa M}.
Note that the monopole here also has been observed in the surface Dirac cone or the Fermi arc state\cite{Li R}.
The simplest Hamiltonian describe the Dirac or Weyl semimetal reads $H=\pm(\hbar v_{F}({\bf k}\mp \alpha)\cdot{\pmb \sigma}+\tau_{z}\beta)$,
where $\tau_{z}$ denote the pseudospin degree of freedom, $\alpha$ and $\beta$ are the terms related to the time-reversal invariance and inversion symmetry,
respectively,
i.e.,
$\alpha$ stands the distance between the K-point to the Weyl nodes in momentum space\cite{Lv M,Chang H R2}
which is an important feature of the Weyl semimetal from teh Dirac semimetal.
While for the two-dimension Dirac system with nonzero Dirac-mass (we here take the monolayer silicene and MoS$_{2}$ as example),
the $\beta$-term (i.e., the pseudospin index-dependent term) is also needed for the Berry curvature and the related anomalous effects\cite{Wu C HX}.
For silicene in low-energy Dirac tight-binding model,
the Hamiltonian of silicene reads
which reads\cite{Wu C HX,Wu C H3,Wu C H1,Wu C H4,Wu C H7,XX,Wu C H_3}
\begin{equation} 
\begin{aligned}
H=&\hbar v_{F}(\eta\tau_{x}k_{x}+\tau_{y}k_{y})+\eta\lambda_{{\rm SOC}}\tau_{z}\sigma_{z}+a\lambda_{R_{2}}\eta\tau_{z}(k_{y}\sigma_{x}-k_{x}\sigma_{y})\\
&-\frac{\overline{\Delta}}{2}E_{\perp}\tau_{z}+\frac{\lambda_{R_{1}}}{2}(\eta\sigma_{y}\tau_{x}-\sigma_{x}\tau_{y})+M_{s}s_{z}
-\eta\tau_{z}\hbar v_{F}^{2}\frac{\mathcal{A}}{\Omega}+\mu,
\end{aligned}
\end{equation}
where 
$E_{\perp}$ is the perpendicularly applied electric field, 
$a=3.86$ is the lattice constant,
$\mu$ is the chemical potential,
$\overline{\Delta}=0.46$ \AA\ is the buckled distance between the upper sublattice and lower sublattice,
$\sigma_{z}$ and $\tau_{z}$ are the spin and sublattice (pseudospin) degrees of freedom, respectively.
$\eta=\pm 1$ for K and K' valley, respectively.
$M_{s}$ is the spin-dependent exchange field. 
$\lambda_{SOC}=3.9$ meV is the strength of intrinsic spin-orbit coupling (SOC) and $\lambda_{R_{2}}=0.7$ meV is the intrinsic Rashba coupling
which is a next-nearest-neightbor (NNN) hopping term and breaks the lattice inversion symmetry.
$\lambda_{R_{1}}$ is the electric field-induced nearest-neighbor (NN) Rashba coupling which has been found that linear with the applied electric field
in our previous works\cite{Wu C H1}: $\lambda_{R_{1}}=0.012E_{\perp}$.
The Dirac-mass and the corresponding quasienergy spectrum (obtained throught the diagonalization procedure) are\cite{Wu C HX,XX,Wu C H_3}
\begin{equation} 
\begin{aligned}
&m_{D}^{\eta,s_{z}}=\eta\sqrt{\lambda_{{\rm SOC}}^{2}+a^{2}\lambda^{2}_{R_{2}}k^{2}}s_{z}\tau_{z}-\frac{\overline{\Delta}}{2}E_{\perp}\tau_{z}+M_{s}s_{z},\\
&\varepsilon=s\sqrt{a^{2}\lambda^{2}_{R_{2}}k^{2}+(\sqrt{\hbar^{2}v_{F}^{2}{\bf k}^{2}
+(\eta\lambda_{{\rm SOC}}s_{z}\tau_{z}-\frac{\overline{\Delta}}{2}E_{\perp}\tau_{z} )^{2}}+M_{s}s_{z}+s\mu)^{2}},
\end{aligned}
\end{equation}
respectively, 
where $s=\pm 1$ is the electron/hole index.
The Dirac-mass here is related to the band gap by the relation $\Delta=2|m_{D}|$.
For the monolayer MoS$_{2}$ which also has the outstanding properties about the
unconventional Quantum Hall effect and spin Hall effect like the silicene,
and with a higher buckled structure (angle $\theta=40.7^{\text{o}}$) than silicene,
the low-energy Dirac Hamiltonian reads\cite{Scholz A,Rostami H}
\begin{equation} 
\begin{aligned}
H=\frac{\hbar^{2}k^{2}}{4m_{0}}(0.43+2.21 s_{z})+m_{D}s_{z}+s_{z}\eta_{z}\lambda_{SOC}\delta(s_{z},-1)+t_{0}a{\rm cos}\theta{\bf k}\cdot{\pmb \sigma},
\end{aligned}
\end{equation}
where $m_{0}$ is the free electron mass, and $\lambda_{SOC}=0.08$ eV, $t_{0}=1.68$ eV, $a=2.43$ \AA\ .
We show in Fig.1(a) the band structure of the silicene with nonzero band gap $\Delta=2|m_{D}|$
and $M_{s}=\lambda_{SOC}=0.0039$ eV.
Due to th existence of exchange field, we can easily to see that the two conduction bands are touching with each orther in K-point 
while the two valence bands are divided (see also our previous works\cite{Wu C H3,Wu C H2,Wu C HX}).
For the MoS$_{2}$ (Fig.1(b) and (c)), we obtained the similar result with Ref.\cite{Scholz A}:
The spin-splitted conduction bands are degenerate with each other while the spin-splitted valence bands are not.
We show in Fig.1 only the band structure of the valley K.
Although the inversion symmetry is broken either by the buckled structure of the Rashba-coupling,
the time-reversal invariance is preserved except we apply the off-resonance circularly
polarized light\cite{Wu C HX,Wu C H5,Wu C H2} or the magnetic field in perpendicular direction,
thus the spin-momentum
locking can be observed, and the orbital magnetic moment as well as the energy shift are opposite in two valleys.
The spin-momentum
locking is the intrinsic feature of the topological insulator, and can also be found in the three-dimension Dirac or Weyl semimetal,

\section{Plasmon dispersion}

In long-wavelength limit (and low-energy), 
the quasiparticle excitation has support the types of Fermions in condensed matter system\cite{Bradlyn B,Lv B Q,Ezawa M2,Wang D}.
In such limit, both the one-\cite{Sarma S D}, two-, and three-dimension electron system own a quantum (nonlocal) plasmon dispersion;
beyond such limit (i.e., with finite ${\bf q}$), the monolayer silicene (or graphene) is still quantum (i.e., $\hbar$-dependent),
but for the bilayer silicene (or graphene) and the normal 2DEG which with parabolic energy dispersion,
the plasmon dispersion are classical except carry out the quantum correction to the ${\bf q}^{3/2}$ order.

As we have discussed\cite{Wu C H3,XX,Wu C H_3}, the plasmon dispersion of silicene in long-wavelength limit reads
\begin{equation} 
\begin{aligned}
\omega_{p}=v_{F}\sqrt{g_{s}g_{v}}\sqrt{\frac{e^{2}{\bf q}E_{F}}{2\epsilon\epsilon^{*}\hbar v_{F}}\left[2-\frac{( m_{D}^{{\rm max}})^{2}+( E_{F}^{{\rm min}})^{2}}{E_{F}^{2}}\right]},\ {\rm for}\ 
E_{F}> |m_{D}|^{{\rm max}},\\
\omega_{p}=v_{F}\sqrt{g_{s}g_{v}}\sqrt{\frac{e^{2}{\bf q}E_{F}}{\epsilon\epsilon^{*}\hbar v_{F}}\left[1-\frac{( m_{D}^{{\rm min}})^{2}}{E_{F}^{2}}\right]},\ {\rm for}\ 
 |m_{D}|^{{\rm max}}>E_{F}> |m_{D}|^{{\rm min}},
\end{aligned}
\end{equation}
where the value of Dirac-mass here can be controlled by modifying the indices $\eta$ or $s_{z}$.
Here the maximum and minimum band gap-dependent Dirac mass can be obtained through the Eq.(2):
\begin{equation} 
\begin{aligned}
|m_{D}|^{{\rm max}}=
|-\sqrt{\lambda_{{\rm SOC}}^{2}+a^{2}\lambda^{2}_{R_{2}}k^{2}}-\frac{\overline{\Delta}}{2}E_{\perp}-M_{s}s_{z}|,\\
|m_{D}|^{{\rm min}}=
|\sqrt{\lambda_{{\rm SOC}}^{2}+a^{2}\lambda^{2}_{R_{2}}k^{2}}-\frac{\overline{\Delta}}{2}E_{\perp}+M_{s}s_{z}|,\\
\end{aligned}
\end{equation}
where we restrict the $M_{s}=0.0039$ eV here as usual and apply $\lambda_{R_{2}}=0.012E_{\perp}$.
Thus we know that the gap difference between the maximum gap and minimum gap 
is as large as
$|4\sqrt{\lambda_{{\rm SOC}}^{2}+a^{2}\lambda^{2}_{R_{2}}k^{2}}+4M_{s}s_{z}|$.
The factor $v_{F}$ in the above expressions of the plasmon dispersion 
is in fact required by the long-wavelength plasmon dispersion in all dimensions\cite{Sarma S D,Stern F}.
$E_{F}=\gamma{\bf k}_{F}=\gamma\sqrt{\pi n}$ with $n$ the charge carrier density per unit two-dimension volume
and ${\bf k}_{F}^{(2)}=\sqrt{4\pi n/g_{s}g_{v}}$,
through this we can see that $\omega_{p}\propto n^{1/4}$ in monolayer silicene but $\omega_{p}\propto n^{1/2}$ in bilayer silicene.
While for the Weyl semimetal,
the Fermi wavevector is ${\bf k}_{F}^{(3)}=\sqrt[3]{6\pi^{2} n^{*}/g_{s}g_{v}}$,
with $n^{*}=n\pm \frac{e^{2}e{\bf E}\cdot{\bf B}}{4\pi^{2}\hbar^{2}c}$
where the later term is due to the chiral anomaly.
In such case, due to the existence of midgap Landau level, the Adler-Bell-Jackiew anomaly occur\cite{Ezawa M}
accompanied with the pumped charges between two Weyl nodes.
For a typical carrier density $n=1\times 10^{12}$ cm$^{-2}$, the Fermi wavevectors can be obtained as
${\bf k}_{F}^{(2)}=\sqrt{n\pi}=1.8\times 10^{8}$ m$^{-1}$ and
${\bf k}_{F}^{(3)}=\sqrt[3]{6\pi^{2} n/g_{s}g_{v}}=2.1\times 10^{5}$ m$^{-1}$ for the undoped case (intrinsic).

For the parabolic energy dispersion systems (like the bilayer islicene or the few-layer black phosphorus), the long-wavelength behavior of the plasmon model 
can be obtained by solving
\begin{equation} 
\begin{aligned}
\epsilon({\bf q},\omega)\approx 1-g_{s}g_{v}\frac{e^{2}E_{F}{\bf q}}{\epsilon\epsilon^{*}\omega_{p}^{2}}=0,
\end{aligned}
\end{equation}
where we simpliy view the unit area of the two-dimension sheet as $A=1$,
and then the plasmon dispersion (in gapped case) can be written as
\begin{equation} 
\begin{aligned}
\omega_{p}=\sqrt{g_{s}g_{v}}\sqrt{\frac{e^{2}{\bf q}E_{F}}{2\epsilon\epsilon^{*}}\left[2-\frac{( m_{D}^{{\rm max}})^{2}+( m_{D}^{{\rm min}})^{2}}{E_{F}^{2}}\right]},\ {\rm for}\ 
E_{F}> |m_{D}|^{{\rm max}},\\
\omega_{p}=\sqrt{g_{s}g_{v}}\sqrt{\frac{e^{2}{\bf q}E_{F}}{\epsilon\epsilon^{*}}\left[1-\frac{( m_{D}^{{\rm min}})^{2}}{E_{F}^{2}}\right]},\ {\rm for}\ 
 |m_{D}|^{{\rm max}}>E_{F}> |m_{D}|^{{\rm min}},
\end{aligned}
\end{equation}
where $E_{F}={\bf k}_{F}^{2}/2m$ here with $m$ the mass related to the intralayer or interlayer hopping.
through this we can see that $\omega_{p}\propto n^{1/2}$ in bilayer silicene and 2DEG,
For the Weyl semimetal,
the plasmon frequency in long-wavelength limit can be approximately solved from
\begin{equation} 
\begin{aligned}
{\rm Re}\epsilon(\omega)=\epsilon\epsilon^{*}-v^{2}_{F}\sqrt{\frac{e^{2}}{\hbar v_{F}\epsilon\epsilon^{*}}}(g_{s}g_{v}\frac{32\pi}{3})^{1/3}n^{2/3}\frac{1}{\omega^{2}}.
\end{aligned}
\end{equation}
The ${\rm Re}\epsilon(\omega)$ 
in fact has a logarithmic dependence on the frequency,
with\cite{Lv M}
\begin{equation} 
\begin{aligned}
\epsilon^{*}=1+\frac{2\pi e^{2}}{\epsilon\epsilon^{*}}D_{F}\left[\frac{{\bf q}^{2}}{12{\bf k}_{F}^{2}}{\rm ln}\left|\frac{\frac{\Lambda^{2}}{{\bf k}_{F}^{2}}}
{\frac{{\bf q}^{2}}{{\bf k}_{F}^{2}}-\frac{\omega^{2}}{\mu^{2}}}\right|
+i\frac{\pi}{3}\frac{{\bf q}^{2}}{4{\bf k}_{F}^{2}}\theta(\frac{\omega}{2\mu}-\frac{{\bf q}}{2{\bf k}_{F}})\right],
\end{aligned}
\end{equation}
the latter term within the bracket (imaginary) also appear in the dielectric function of the Kane Fermions\cite{Orlita M},
where the latter term becomes $i{\rm sgn}\omega$ due to the optical limit (${\bf q}\rightarrow 0$). 
For the case of gapped Weyl system, 
the factor $\hbar$ vanishes due to the classical limit in the first order,
and the plasmon frequency can then be obtained as
\begin{equation} 
\begin{aligned}
\omega_{p}=v_{F}\sqrt{\frac{e^{2}}{2\epsilon\epsilon^{*}v_{F}^{2}}(\frac{32\pi}{3})^{1/6}n^{1/3}\sqrt[3]
{[2-\frac{( m_{D}^{{\rm max}})^{3}+( m_{D}^{{\rm min}})^{3}}{E_{F}^{3}}]}},\ {\rm for}\ 
E_{F}> |m_{D}|^{{\rm max}},\\
\omega_{p}=v_{F}\sqrt{\frac{e^{2}}{2\epsilon\epsilon^{*}v_{F}^{2}}(\frac{32\pi}{3})^{1/6}n^{1/3}\sqrt[3]
{[1-\frac{( m_{D}^{{\rm min}})^{3}}{E_{F}^{3}}]}},\ {\rm for}\ 
 |m_{D}|^{{\rm max}}>E_{F}> |m_{D}|^{{\rm min}},
\end{aligned}
\end{equation}
which is similar to the results of Refs.\cite{Lv M,Sarma S D,Zhou J} which are for the gapless case.

\section{Relaxation time and the dc conductivity}

In the presence of the short-range impurity (with $\delta$-term, i.e., the scattering reads $V_{s}(r)=V({\bf q})\delta(r-r_{{\rm imp}})$) 
in the noninteracting approximation
(which in fact has high accuracy in low-temperature case),
the momentum relaxation rate (impurity scattering rate) can be obtained by the first Born approximation,
\begin{equation} 
\begin{aligned}
\frac{1}{\tau_{2}}=-V^{2}_{2}n_{i}{\rm Im}\int\frac{d^{2}q}{(2\pi)^{2}}G({\bf q},\omega)
\sim V^{2}_{2}n_{i}\frac{4\pi\varepsilon}{v_{F}^{2}},
\end{aligned}
\end{equation}
where
the impurities scattering potential after the Fourier transformation is $V_{2}=\frac{2\pi e^{2}}
{\epsilon_{0}\epsilon\sqrt{{\bf q}^{2}+{\bf k}_{s}^{2}}}$
with the screening wave vector ${\bf k}_{s}=2\pi e^{2}\Pi({\bf q},\omega)/(\epsilon_{0}\epsilon)$ which is polarization-dependent.
$G({\bf q},\omega)$ is the the retarded lattice Green's function
\begin{equation} 
\begin{aligned}
G({\bf q},\omega)=[\omega-H({\bf q})-\Sigma(\omega)]^{-1},
\end{aligned}
\end{equation}
where $\Sigma(\omega)$ is the self-energy matrix
and indeed can also be omitted here due to the noninteracting assumation.
Here the self-energy as well as the vertex correction origin from the electron interaction can in fact be ignored due to the low-energy
assumation,
in which case that the disorder is dominates in low-energy transport but not the electron interaction\cite{Burkov A A}.
The denisty of state in Fermi level here is $D_{F}=\sqrt{g_{s}g_{v}\frac{n}{\pi\gamma^{2}}}$.
Similarly, the relaxation rate in three dimension undoped Dirac or Weyl system can be obtained as
\begin{equation} 
\begin{aligned}
\frac{1}{\tau_{3}}=-V_{3}^{2}n_{i}{\rm Im}\int\frac{d^{3}q}{(2\pi)^{3}}G({\bf q},\omega)=2\pi V_{3}^{2}n_{i}D_{F}^{3}
\sim V_{3}^{2}n_{i}\frac{8\pi^{2}\varepsilon^{2}}{v^{3}_{F}}
\sim V_{3}^{2}n_{i}\frac{8\pi^{2}9g_{s}g_{v}n_{i}^{2}}{v^{3}_{F}},
\end{aligned}
\end{equation}
where
the impurities scattering potential after the Fourier transformation is $V_{3}=\frac{4\pi e^{2}}
{\epsilon_{0}\epsilon({\bf q}^{2}+{\bf k}_{s}^{2})}$
with the screening wave vector ${\bf k}_{s}=2\pi e^{2}\Pi({\bf q},\omega)/(\epsilon_{0}\epsilon)$ which is polarization-dependent.
The denisty of state in Fermi level here becomes $D_{F}=\sqrt[3]{9g_{s}g_{v}\frac{n^{2}}{2\pi^{2}\gamma^{3}}}$
($\gamma\sim v_{F}$ here).
Here we note that, due to the large charge density,
we have $\frac{1}{\tau_{2}}\sim n^{2}$ and $\frac{1}{\tau_{3}}\sim n^{3}$. 

For the long-range impurity (with long-range Coulomb interaction),
the screening wave vector (${\bf k}_{s}$) in the above screened Coulomb interaction $V_{2}$ or $V_{3}$ can be replaced by the Thomas-Fermi wavevector:
\begin{equation} 
\begin{aligned}
{\bf q}_{TF}=g_{s}g_{v}r_{s}{\bf k}_{F}=g_{s}g_{v}\frac{e^{2}}{\hbar \epsilon\epsilon^{*}\gamma}{\bf k}_{F}\\
=g_{s}g_{v}\frac{e^{2}}{\hbar \epsilon\epsilon^{*}}{\bf k}_{F}\frac{\pi\Pi({\bf q},0)\hbar}{2{\bf k_{F}}}\\
=g_{s}g_{v}\frac{e^{2}}{\hbar \epsilon\epsilon^{*}}\frac{\pi\Pi({\bf q},0)\hbar}{2}
\end{aligned}
\end{equation}
for the two-dimension monolayer Dirac system, 
\begin{equation} 
\begin{aligned}
{\bf q}_{TF}=g_{s}g_{v}r_{s}{\bf k}_{F}=g_{s}g_{v}\frac{e^{2}m^{*}}{{\bf k}_{F}\epsilon\epsilon^{*}\hbar^{2}}{\bf k}_{F}\\
=g_{s}g_{v}\frac{e^{2}m^{*}}{\epsilon\epsilon^{*}\hbar^{2}},\\
\end{aligned}
\end{equation}
for the two-dimension bilayer Dirac system, 
and 
${\bf q}^{2}_{TF}=4\pi e^{2}D_{F}^{3}/(\epsilon\epsilon^{*}E_{F})$ for three-dimension Weyl system\cite{Burkov A A},
here we note that the quasienergy for three-dimension Weyl system has $\varepsilon\sim 3\sqrt{g_{s}g_{v}}n$,
and both of them are dependent on the carrier density.
Furthermore, at long-wavelength limit, the Thomas-Fermi wavevector as well as the ${\bf k}_{F}$ for the two-dimension Dirac system is 
proportional to the $D_{F}$.
Then the resulting relaxation rate for the elastic scattering in two-dimension Dirac system reads
\begin{equation} 
\begin{aligned}
\frac{1}{\tau_{2}}=&\frac{2\pi n_{i}}{\hbar}|U({\bf q})|^{2}(1-{\rm cos}\theta)\delta(E_{k}-E_{k'})\\
\end{aligned}
\end{equation}
where $\theta$ is the angle between the wave vectors before and after scattering,
$U({\bf q})=\frac{e^{2}}{2\epsilon\epsilon^{*}\sqrt{{\bf q}^{2}+{\bf k}_{s}^{2}}}$ is the Fourier transform of the 
$U(r)=\frac{e^{2}e^{-{\bf k}_{s}}}{4\pi\epsilon\epsilon^{*}}$.
The factor $(1-{\rm cos}\theta)$ only exist in the presence of the dominating elastic backscattering
and the $\delta$-impurity-term here can be preserved for long-range impurity as a result of the low-temperature.
It appear here as a standard factor\cite{Burkov A A,Stauber T}
and the Fermi golden rule is used here.
The relaxation time is inverse proportion to the conductivity since it's  proportion to the impurity concentration $n_{i}$.
The scattering wave vector reads
${\bf q}={\bf k}-{\bf k}'=2k{\rm sin}(\theta/2)$\cite{Shakouri K,Vargiamidis V},
and here $k\sim \frac{\varepsilon}{\hbar v_{F}}\sim n_{i}^{1/2}$.
For such case (in zero temperature limit), the static (dc) conductivity is related to the impurity concentration and the mean-free path:
$\sigma_{dc}\approx \frac{e^{2}}{h}n_{i}^{1/2}\ell_{2}{\bf k}_{F}\approx \frac{e^{2}}{h}{\bf k}_{F}$ in two-dimension Dirac system,
$\sigma_{dc}\approx \frac{e^{2}}{h}n_{i}^{1/3}\ell_{3}{\bf k}_{F}\approx \frac{e^{2}}{h}{\bf k}_{F}$ in three-dimension Weyl system,
here the Fermi wavevector ${\bf k}_{F}$ for each system has been presented above.
For the dominating interband elastic scattering in gapless case,
the above expression becomes\cite{Vargiamidis V}
\begin{equation} 
\begin{aligned}
\frac{1}{\tau_{2}}
=&\frac{n_{i}}{4\pi\hbar}\frac{\varepsilon}{\hbar^{2}v_{F}^{2}}\int^{\pi}_{0} (1-{\rm cos}^{2}\theta)|U({\bf q})|^{2}d\theta,
\end{aligned}
\end{equation}
For the case of finite temperature $T$ together with the electron interactions
and the unscreened Coulomb interaction, 
the scattering rate becomes $1/\tau\sim \alpha^{2}{\rm max}[\varepsilon,T]$\cite{Burkov A A},
i.e., it's temperature- and quasienergy-dependent (in fact it's also chemical potential- (doping) and frequency-dependent)
except at the zero energy Dirac-point with large compressibility (and thus with the minimal conductivity) and electronic phase separation\cite{Guinea F} 
and supressed orthogonality catastrophe\cite{Hentschel M}.
The minimal conductivity here is independent of the temperature and doping (chemical potential)\cite{Stauber T}
and can be obtained by the previous expression of the dc conductivity at zero temperature $\sigma_{dc}=\frac{e^{2}v_{F}^{2}}{h}D_{F}\tau_{2}=
\frac{e^{2}v_{F}}{h}\sqrt{g_{s}g_{v}\frac{n_{i}}{\pi}}\tau_{2}$,
which reads $\sigma_{dc}^{{\rm min}}=\frac{e^{2}v_{F}}{h}\tau_{2}{\rm max}[{\bf k}_{F},\pi\alpha\sqrt{n}]$
where $\alpha \lesssim 1$ is a dimensionless constant,
note that here $n\neq n_{i}$ is the carrier density.

For the three-dimension Weyl semimetal in the presence of the Donor impurities, the transport collision rate due to impurity scattering is
\begin{equation} 
\begin{aligned}
\frac{1}{\tau_{3}}
=\frac{\pi n_{i}D_{F}^{3}}{2}\int^{\pi}_{0} {\rm sin}\theta |V({\bf q})|^{2}(1-{\rm cos}^{2}\theta)d\theta.
\end{aligned}
\end{equation}

In Fig.2, we present the longitudinal conductivity in two-dimension (gapped) Dirac system with different Dirac-mass in the noninteracting approximation,
which is $\sigma_{xx}=ie^{2}\omega\Pi({\bf q},\omega)/{\bf q}^{2}$.
We see that the imaginary part of the longitudinal conductivity is smaller than zero,
thus we only focus on the real conductivity in the following.
For the real part in long-wavelength limit and in gapless case (Fig.3(b)),
we obtain the well known frequency-independent conductivity $\sigma_{xx}=e^{2}/4\hbar=0.25$\cite{Ludwig A W W},
while for the gapped case, the longitudinal conductivity close to the $4e^{2}/h\approx 0.6336$ is obtained.
For the three-dimension Dirac or Weyl gapped system,
the real part of the longitudinal conductivity reads
\begin{equation} 
\begin{aligned}
{\rm Re}\sigma_{xx}=-\frac{e^{2}v_{F}}{\omega}{\rm Im}\Pi({\bf q},\omega),
\end{aligned}
\end{equation}
in the optical limit (${\bf q}\rightarrow 0$),
the above expression becomes the optical conductivity which can be verified by the Kubo formula\cite{Orlita M,Thakur A,Sodemann I}.
We present in Fig.3 the longitudinal conductivity in three-dimension Dirac or Weyl gapped system,
where we shown in left column corresponds to the case of small chemical potential $\mu<m_{D}$ while the second column corresponds to
the case of large chemical potential $\mu>m_{D}$.
We obtained different result compared to the Fig.2, e.g., the longitudinal conductivity in long-wavelength limit vanishes 
no matter how large the chemical potential is.
Further, the longitudinal conductivity shown in Fig.3 has a very small value in the optical limit,
and nonzero only in the regime $\omega>{\bf q}$ and $\omega>2\mu$,
which is agreed with the result of Ref.\cite{Thakur A}
where the optical conductivity is obtained as ${\rm Re}\sigma_{xx}\approx e^{2}\omega/24\pi\hbar v_{F}$
and equals $0.013$ in static case ($\omega=0$).

\section{Friedel oscillation}

The electron-density-deviation induced by the impurity is $\delta n=\int\frac{d^{2}q}{(2\pi)^{2}}(1-\epsilon({\bf q}))$.
Within the RPA in static limit, the screened potential of the charged impurities which determined by the charge density as well as the position reads
\begin{equation} 
\begin{aligned}
\phi(r)=&\frac{Ze}{\alpha\epsilon\epsilon^{*}}\int\frac{d^{2}q}{(2\pi)^{2}}\frac{2\pi\alpha e^{i{\bf q}\cdot{\bf r}}}{{\bf q}\epsilon({\bf q})}\\
=&\frac{Ze}{\epsilon\epsilon^{*}}\int\frac{d^{2}q}{2\pi}\frac{e^{i{\bf q}\cdot{\bf r}}}{{\bf q}\epsilon({\bf q})}\\
=&\frac{Ze}{\epsilon\epsilon^{*}}\int\frac{d^{2}q}{2\pi}\frac{e^{i{\bf q}\cdot{\bf r}}}{{\bf q}-(2\pi e^{2}/\epsilon\epsilon^{*})\Pi({\bf q})}\\
=&\frac{Ze}{\epsilon\epsilon^{*}}\int^{\infty}_{0}\frac{qdq}{{\bf q}-(2\pi e^{2}/\epsilon\epsilon^{*})\Pi({\bf q})}\int^{\pi}_{0}{\rm sin}\theta 
e^{iqr{\rm cos}\theta}d\theta\\
=&\frac{Ze}{\epsilon\epsilon^{*}}\int^{\infty}_{0}\frac{qdq}{{\bf q}-(2\pi e^{2}/\epsilon\epsilon^{*})\Pi({\bf q})}\frac{i(e^{-iqr}-e^{iqr})}{qr}\\
=&\frac{Ze}{\epsilon\epsilon^{*}}\int^{\infty}_{0}\frac{qdq}{{\bf q}-(2\pi e^{2}/\epsilon\epsilon^{*})\Pi({\bf q})}J_{0}(qr),\\
\end{aligned}
\end{equation}
where $\alpha=e^{2}/\hbar v_{F}\epsilon$ is the fine structure constant
which has a close definition with the $r_{s}$:
The ratio of the Coulomb interaction to the kinetic energy.
$J_{0}(qr)$ is the zeroth Bessel function of the first kind.
$Ze$ is the charge of the impurity.
The static polarization here reads\cite{Wu C H3,XX,Wu C H_3,Tabert C J}
\begin{equation} 
\begin{aligned}
\Pi({\bf q})=-g_{s}g_{v}\frac{2e^{2}\mu}{\epsilon\epsilon^{*}\hbar^{2}v_{F}^{2}}
\left[\frac{ m_{D}}{2\mu}+\frac{\hbar^{2} v_{F}^{2}{\bf q}^{2}-4 m_{D}^{2}}{4\hbar v_{F}{\bf q}\mu}{\rm arcsin}\sqrt{\frac{\hbar^{2}v_{F}^{2}{\bf q}^{2}}{\hbar^{2}v_{F}^{2}{\bf q}^{2}+4 m_{D}^{2}}}\right]
\end{aligned}
\end{equation}
for $0<\mu< m_{D}$, and
\begin{equation} 
\begin{aligned}
\Pi({\bf q})=-g_{s}g_{v}\frac{2e^{2}\mu}{2\pi\epsilon_{0}\epsilon \hbar^{2} v_{F}^{2}}
\left[1-\Theta({\bf q}-2{\bf k}_{F})\left(\frac{\hbar^{2} v_{F}^{2}\sqrt{{\bf q}^{2}-4{\bf k}_{F}^{2}}}{2\hbar v_{F}{\bf q}}-\frac{\hbar^{2}v_{F}^{2}{\bf q}^{2}-4 m_{D}^{2}}{4\mu \hbar v_{F}{\bf q}}{\rm arctan}\frac{\hbar v_{F}\sqrt{{\bf q}^{2}-4{\bf k}^{2}_{F}}}{2\mu}\right)\right]
\end{aligned}
\end{equation}
for $\mu> m_{D}$.
Through the first four lines of the above expression of the polarization 
at large wave vector (momentum) ${\bf q}$ and short distance, the screened static Coulomb potential
$\phi(r)\sim \frac{Ze}{\epsilon\epsilon^{*}r}$,
which implies a simply logarithmic decay with the increase of distance,
while in long-distance regime ($r\gg {\bf k}_{F}^{-1}$), the screened potential decrease more faster as $\sim r^{3}$\cite{Pyatkovskiy P K}
since the intravalley backscattering is suppressed and that's similar to the
decay of the quasi-particle interference pattern in silicene which is also suppressed by the intravalley scattering 
(corresponds to the long-wavelength interference)
but restored by the elastic intervalley scattering\cite{XX}.
In Fig.4, we show the 
screened potential as a function of distance,
the Friedel oscillation origin from the non-analyticity of the polarization at ${\bf q}=2{\bf k}_{F}$,
and dominates the contribution to screened potential in the case of short-range Coulomb potential and large distance (i.e., small wavevector ${\bf q}$).
The Friedel oscillation
vanishes for the case of $\mu<m_{D}^{{\rm min}}$ (see Fig.4(c)-(d)).
In the first column of panels (a) and (b), we present the results for the case of $m_{D}^{{\rm max}}>0$ eV
and $0<m_{D}^{{\rm min}}<m_{D}^{{\rm max}}$ eV;
in the second column of panels (a) and (b), we present the results for the case of $m_{D}^{{\rm max}}>0$ eV
and $m_{D}^{{\rm min}}=0$ eV,
i.e., the gapless case where we can see that the resulting screened potential decays as $\sim 1/r^{3}$.
In (a) and (c), we set $m_{D}^{{\rm max}}=0.2$ eV, and in (b) and (d) $m_{D}^{{\rm max}}=1.2$ eV.
The first row of the (a) and (b) panels contain only the contribution of the ${\bf q}>2{\bf k}_{F}$ part,
where we can not see the beating effect, and the ;
the second row of the (a) and (b) panels contain both the contributions of the ${\bf q}>2{\bf k}_{F}$ part and ${\bf q}<2{\bf k}_{F}$ part,
where we can easily see the beating effect.
The zero $m_{D}^{{\rm min}}$ leads to continuous first derivative 
of polarization at ${\bf q}=2{\bf k}_{F}^{{\rm max}}$ where ${\bf k}_{F}^{{\rm max}}=\sqrt{\mu^{2}-(m_{D}^{{\rm min}})^{2}}$,
while when $m_{D}^{{\rm max}}=m_{D}^{{\rm min}}=0$, then ${\bf k}_{F}^{{\rm min}}={\bf k}_{F}^{{\rm max}}$
and both of these two Fermi wavevectors have continuous first derivative in ${\bf q}=2{\bf k}_{F}$.
The static polarization has $d\Pi({\bf q})/dq\propto 1/\sqrt{{\bf q}^{2}-4{\bf k}_{F}^{2}}$ for the two-dimension Dirac system\cite{Hwang E H2}
and $d\Pi({\bf q})/dq\propto {\rm ln}|{\bf q}-2{\bf k}_{F}|$ for the three-dimension Weyl system\cite{Lv M}.
The strong decay of the screened potential appear in the Weyl semimetal or the monolayer dice lattice (or the multilayer graphene\cite{Koshino M})
which acts as $\sim r^{-4}$
are not exactly the same,
the reason for the latter ones is due to the strong electronic screening effect of the flat band structure (heavy hole) in the Fermi level.
The flat band here in zero energy is usually not the Dirac or Weyl node line found in three-dimension Dirac or Weyl system
due to the missing of chiral symmetry in these systems.
Such flat band structure can also be found in the Kane Fermion model in two- or three-dimension\cite{Malcolm J D2}.
The interband contribution from the flat band to the conduction band or valence band is cover the whole ${\bf q}\sim\omega$ space\cite{Malcolm J D}. 
In the presence of magnetic impurity, the RKKY interaction which determined by the polarization strength is rised,
and distincted from the electric impurity, the decay of the screened potential $\phi(r)$ is completely contributed from the 
Friedel oscillation but not the Thomas-Fermi contribution.

The Friedel oscillation is dominates in the short-range Coulomb interaction while the Thomas-Fermi contribution is
dominates in the long-range Coulomb interaction\cite{Thakur A2}.
Through the Thomas-Fermi wave vector obtained above,
the screened potential at large distance $r$ due to the Friedel oscillation contribution can be obtained as
\begin{equation} 
\begin{aligned}
\phi(r)=\frac{Ze}{\epsilon\epsilon^{*}}\frac{{\bf q}_{TF}}{(2{\bf k}_{F}+{\bf q}_{TF})^{2}}\frac{{\rm sin}(2{\bf k}_{F}r)}{r^{2}},
\end{aligned}
\end{equation}
for 2DEG (the same as the parabolic system, like the bilayer silicene), and\cite{Lv M}
\begin{equation} 
\begin{aligned}
\phi(r)=\frac{Ze}{\epsilon\epsilon^{*}}\frac{{\bf q}_{TF}^{2}}
{{\bf k}_{F}(4{\bf k}^{2}_{F}+\frac{2}{3}{\rm ln}2{\bf q}_{TF})^{2}}\frac{{\rm sin}(2{\bf k}_{F}r)}{r^{4}},
\end{aligned}
\end{equation}
for the three-dimension Weyl semimetal.
As we mentioned above, the massive (gapped) two-dimension Dirac-system as well as the 2DEG and the parabolic system
decay as $r^{-2}$ for large distance,
for massless (gapless) three-dimension Dirac or Weyl semimetal, the Friedel oscillation decay as $r^{-4}$
due to the chirality effect which supress the back scattering in ${\bf q}=2{\bf k}_{F}$ even in the low-temperature.
While for the parabolic system with both the Rashba-coupling and the Dresselhaus-coupling in unequal strength
(thus with the broken SU(2) symmetry),
the doubly singularities occur\cite{Badalyan S M} due to their effect to the interband transition\cite{Li Z}.

\section{Conclusion}

We discuss the dynamical polarization, plasmon dispersion, relaxation time (or transport time for the large distance case)
as well as the longitudinal conductivity, and the Friedel oscillation of screened potential 
of the two-dimension Dirac and three-dimension Weyl system.
The results, including the Fermi wavevector, Thomas-Fermi wavevector, longitudinal conductivity, and the density of states in Fermi level between different
dimensions are compared.
Some important conclusions are also detaily discussed in the text, including the 
screening character under short or long range Coulomb interaction.
The model for which the calculations based on is the low-energy tight-binding model as described in the Sec.2,
i.e., the results in this paper is for the low-temperature and low-energy case
and it's thus possible to carrying out a logarithmic self-energy correction to the relaxation time as we discussed in the text.
The difference between the longitudinal conductivities in serversal systems is also been discussed.
Our results is helpful to the application of the Dirac or Weyl systems as well as the study on their low-temperature characters.

\end{large}
\renewcommand\refname{References}

\clearpage
Fig.1
\begin{figure}[!ht]
   \centering
 \centering
   \begin{center}
     \includegraphics*[width=1\linewidth]{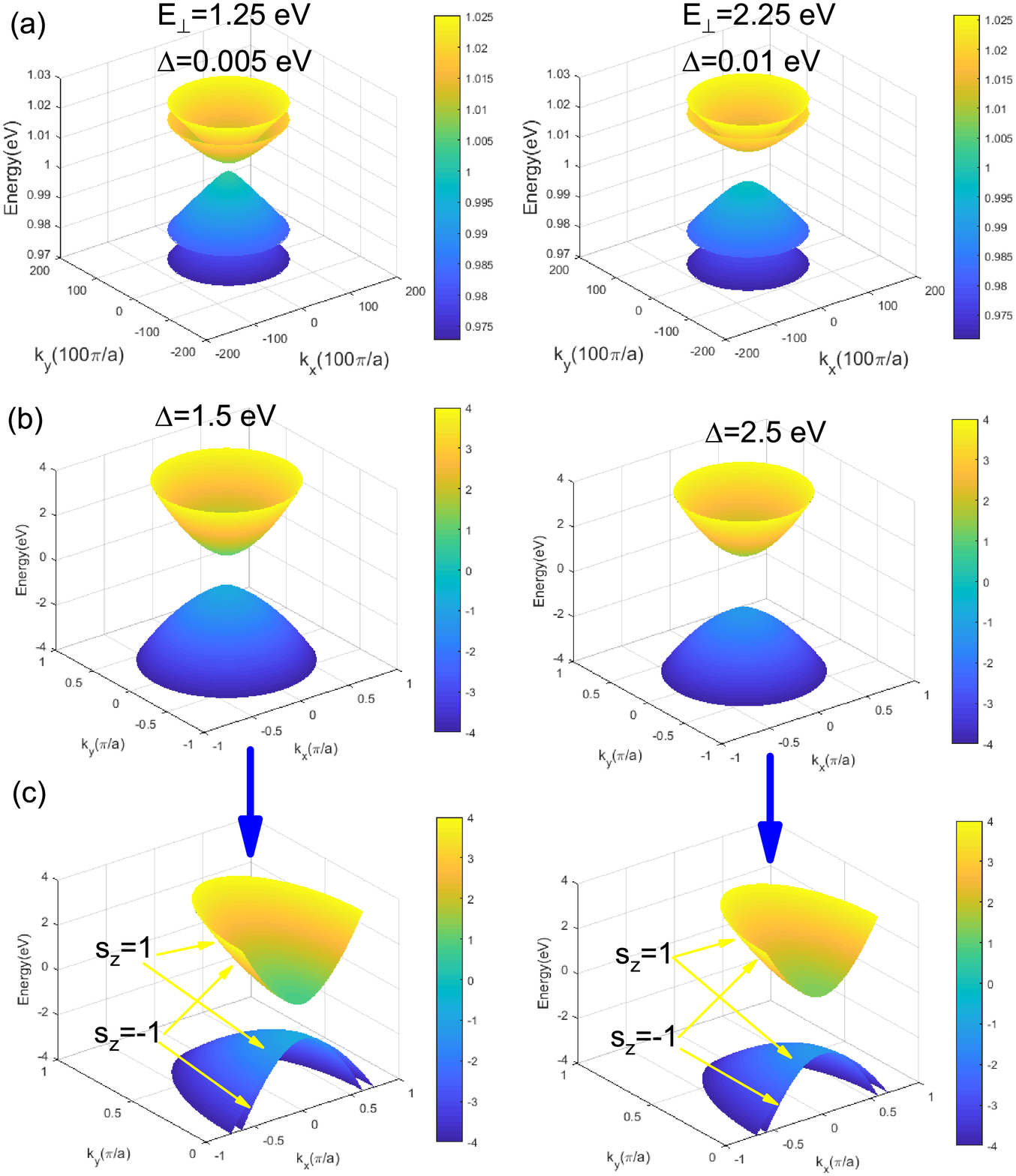}
\caption{(Color online) The low-energy band structure of monolayer silicene (a) and MoS$_{2}$ ((b) and (c)) at valley K ($\eta=1$).
(c) is the profile of (b).
The weak trigonal warping term are ignored here.
}
   \end{center}
\end{figure}

\clearpage
Fig.2
\begin{figure}[!ht]
   \centering
 \centering
   \begin{center}
     \includegraphics*[width=1\linewidth]{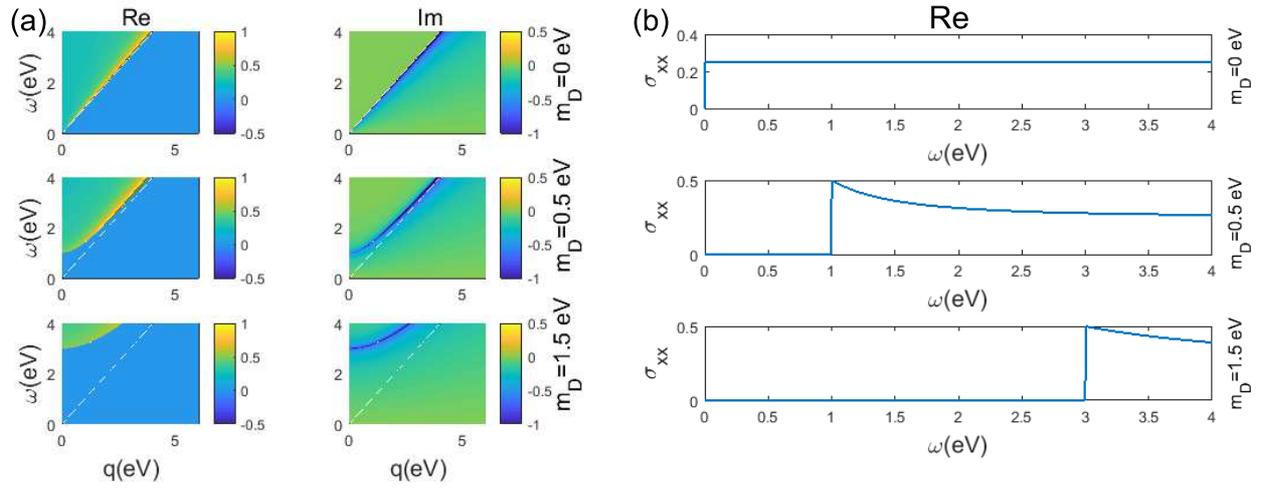}
\caption{(Color online) The longitudinal conductivity of two-dimension Dirac system in ${\bf q}\sim\omega$ space (a),
and in long-wavelength limit (${\bf q}\rightarrow 0$)(b).
We set here $\mu<m_{D}$ (we have present the reason in Ref.\cite{Wu C H_3}).
The corresponding Dirac-mass are indicated in the right-side of each panel.
The spin and valley species are considered here.
}
   \end{center}
\end{figure}
\clearpage
Fig.3
\begin{figure}[!ht]
   \centering
 \centering
   \begin{center}
     \includegraphics*[width=1\linewidth]{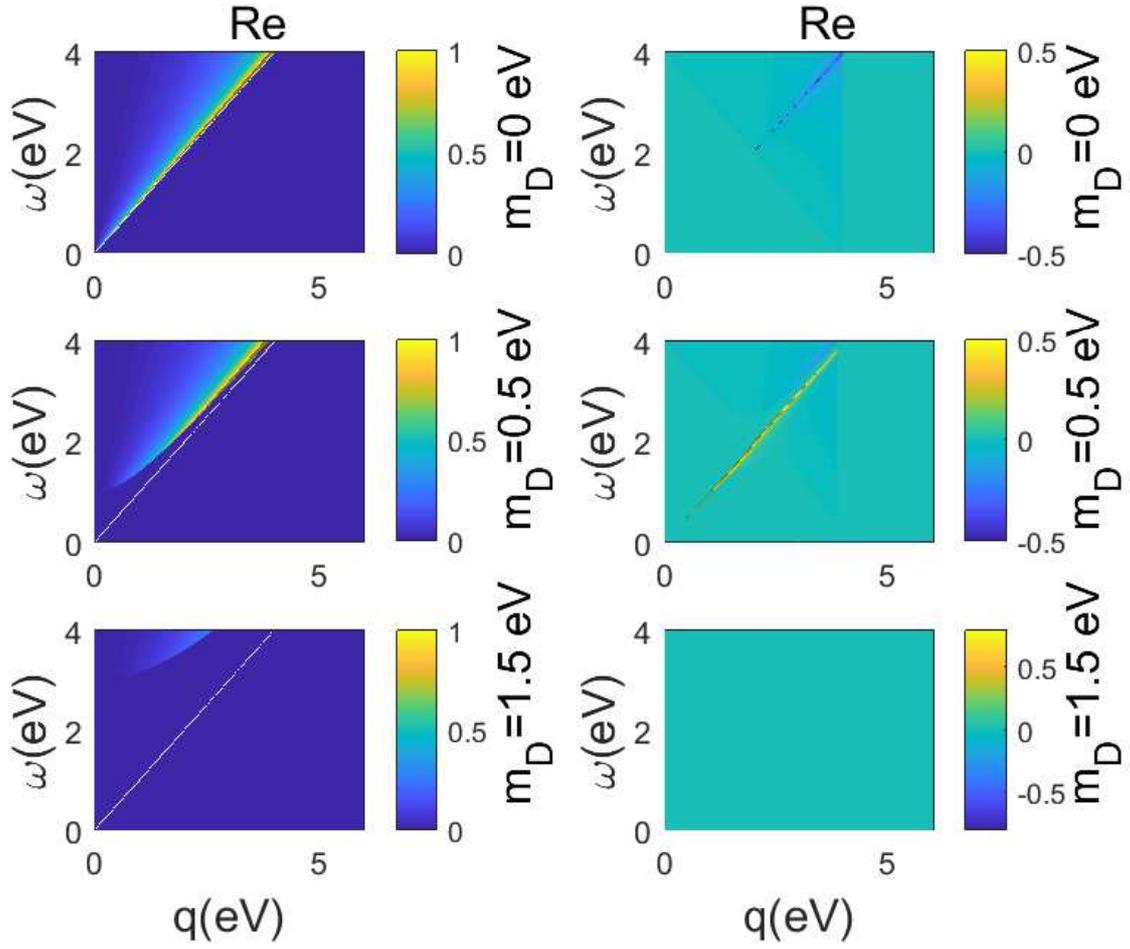}
\caption{(Color online) The longitudinal conductivity of three-dimension Dirac system in ${\bf q}\sim\omega$ space.
The left column is for the case $\mu<m_{D}$
and the second column is for the case of $\mu>m_{D}$.
The corresponding Dirac-mass are indicated in the right-side of each panel.
The spin and valley species are considered here.
}
   \end{center}
\end{figure}

\clearpage
Fig.4
\begin{figure}[!ht]
   \centering
 \centering
   \begin{center}
     \includegraphics*[width=1\linewidth]{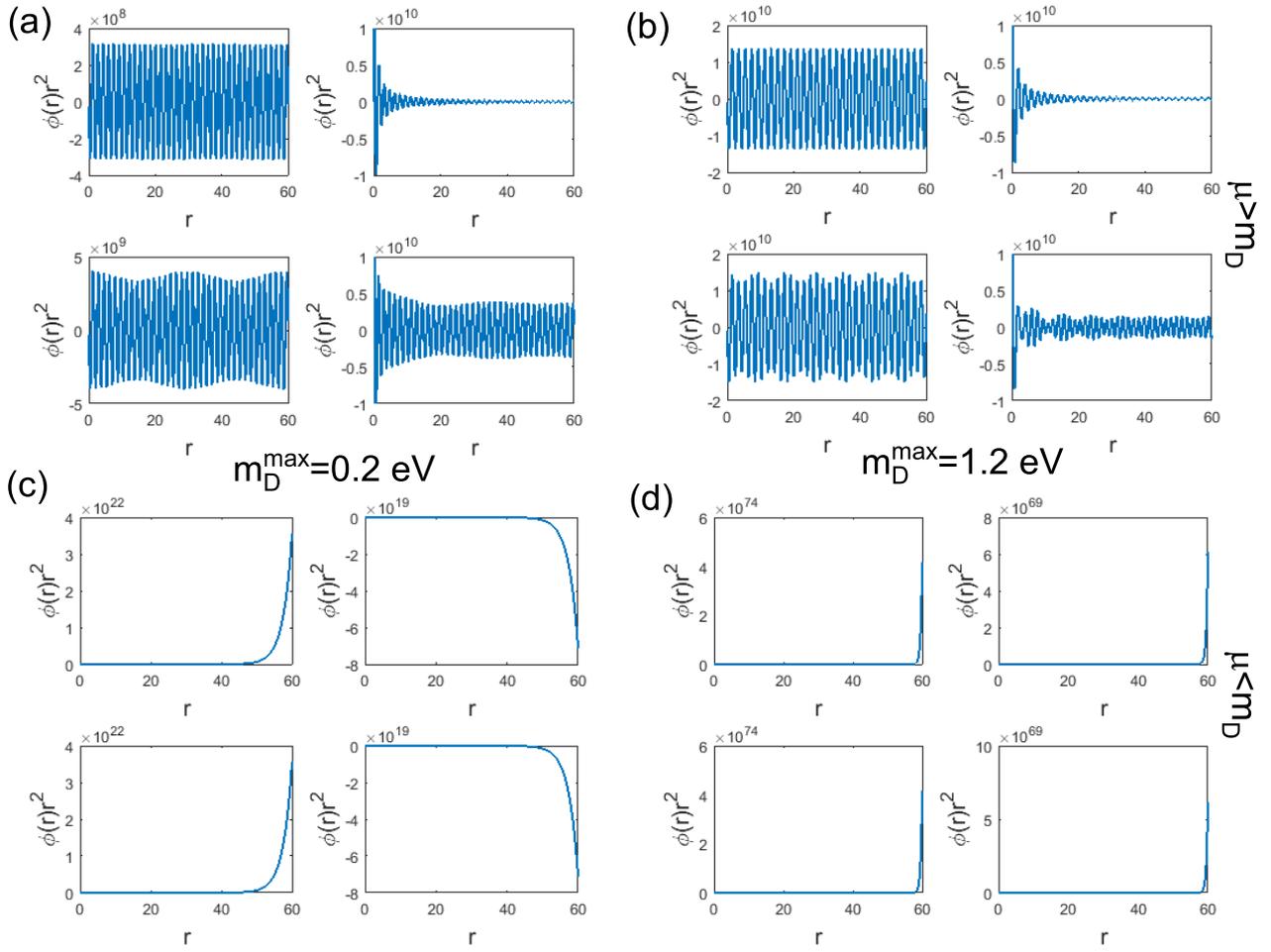}
\caption{(Color online) Friedel oscillation of the screened static Coulomb potential.
In (a) and (b), we set $\mu>m_{D}^{{\rm max}}$,
in (c) and (d), we set $\mu<m_{D}^{{\rm min}}$.
In (a) and (c), we set $m_{D}^{{\rm max}}=0.2$ eV, and in (b) and (d) $m_{D}^{{\rm max}}=1.2$ eV.
The first row of the (a) and (b) panels contain only the contribution of the ${\bf q}>2{\bf k}_{F}$ part,
where we can not see the beating effect;
the second row of the (a) and (b) panels contain both the contributions of the ${\bf q}>2{\bf k}_{F}$ part and ${\bf q}<2{\bf k}_{F}$ part,
where we can easily see the beating effect.
}
   \end{center}
\end{figure}

\end{document}